\documentclass[%
 reprint,
 amsmath,amssymb,
 aps,
]{revtex4-2}

\usepackage{graphicx}% Include figure files
\usepackage{dcolumn}% Align table columns on decimal point
\usepackage{bm}% bold math
\usepackage[utf8]{inputenc}
\usepackage[T1]{fontenc}
\usepackage{mathptmx}
\usepackage[dvipsnames]{xcolor}
\usepackage{multirow}

\begin{document}
\preprint{AIP/123-QED}

\title{Analytical Modeling of Electromagnetic Rotation in Nonreciprocal Media}
% Force line breaks with \\

\author{Swadesh Poddar}
\email{poddarswadesh@gmail.com}
\affiliation{Department of Electrical Engineering, University of Wisconsin-Milwaukee, Milwaukee, Wisconsin 53211, USA}%

\author{Ragib Shakil Rafi}
\email{ragib@ku.edu}
\affiliation{Department of Electrical Engineering \& Computer Science, KS 66045, USA}%

\author{Md. Tanvir Hasan}
\email{tan_vir_bd@yahoo.com}
\affiliation{Department of Electrical and Electronic Engineering (EEE),  
Jashore University of Science and Technology (JUST), Bangladesh}

\date{\today}
            
\begin{abstract}
 Abstract- Reciprocity is a fundamental principle follows the time reversal symmetry of physics. However, many practical applications require breaking time reversal symmetry, hence, are called nonreciprocal. This article aims at discussing time reversal symmetry, developing fundamental building block to achieve nonreciprocity leading to robust analytical model to explain electromagnetic rotation upon
propagation through nonreciprocal medium. Detailed mathematical derivation is presented for Faraday and Kerr rotation in the presence of external bias which breaks time reversal symmetry and leads to achieve nonreciprocal system. We validate our proposed model for conventional conditions and we compute the Faraday and Kerr rotation from a reported article using our proposed mathematical model and observed excellent agreement.
\end{abstract}

\maketitle

\section{\label{sec:level1}Introduction}

\bigskip Physics, the science of nature, in which wave-matter interaction plays a pivotal role paves the foundation of many innovative branches leading to novel applications. In classical electromagnetism a system can be subdivided into two broad categories based on its interaction with electromagnetic wave: reciprocal and non-reciprocal. 

A system is called reciprocal when transmission between transmitter and receiver are identical in both forward and reverse directions. Reciprocity principle in two ports linear circuits can be represented as $V_{1}I_{1}^{'} = V_{2}^{'}I_2$ where voltage $V_1$ applied to first port create current $I_2$ to second port in the same circuit and similarly, voltage $V_{2}{'}$ applied to second branch create current $I_1$ in the first branch \cite{kord,Maxwell}. A two port reciprocal system can be represented in Scattering parameter as $\mathbf{S} = \mathbf{S^T}$. On the other hand nonreciprocity, opposite of reciprocity, is a very important concept at both microwave and optical regime dictates that the fields created by the source at the observation point are different when source and observation points exchange position. A two port nonreciprocal system can be represented as $S_{21}$ $\neq$ $S_{12}$.

In our daily life applications most of the devices such as antenna, electrical circuits, and components are reciprocal. However, reciprocity not only imposes stringent restrictions on how the devices will operate, but also nonreciprocal system offers unique features that will lead to plethora of new applications. Non reciprocity can be achieved by breaking time reversal symmetry. As an example nonreciprocal components such as gyrators, isolators, and circulators, are very important now-a-days because of their novel application oriented features \cite{Nagulu2020}. In 1845, Michael Faraday discovered the first physical relationship between light and magnetism where he demonstrated that a magnetic field in the direction of propagation of a light beam in a transparent medium produces the effects of circular birefringence, which is now known as the Faraday rotation, or the Faraday effect \cite{ferromagnetic:faraday}. Conceptually Faraday effect refers to the specific situation where there is a different index of refraction which correspond to different impedance and phase velocities for left- and
right-handed circularly (L/RHCP) polarized waves that propagate parallel to the external magnetic
field. This LHCP and RHCP wave start to accumulate
a phase difference as they propagate through the medium, therefore, with time and distance the direction of linear polarization
changes. On the other hand, Kerr rotation, discovered by John Kerr in 1875,  is the effect of the difference in phase angle delays resulting from the left and right circularly polarized part of light waves upon normal incidence reflection from a magnetic material. If the material shows time-reversal symmetry, both polarized waves have the same rotation angles; Kerr rotation is nonzero, and the material breaks time-reversal symmetry and becomes nonreciprocal. Both Faraday and Kerr rotations are important metrics to quantify nonreciprocal behavior. However, there is no systematical step by step mathematical approach of the polarization rotation for a specific system. Therefor, detailed theoretical understandings of the polarization rotation are of immense importance.Therefore a robust analytical model been developed to calculate polarization rotation in a conventional anisotrpic environment.

The paper is organized as follows. In Section II, we provide insight to the mechanism  of EM nonreciprocity and various polarization rotation, in Section III we provide details on the step-by-step analytical approach of modelling Faraday and Kerr rotation.

\section{Electromagnetic Non-reciprocity and Polarization Rotation}
Non-reciprocity is the building block of various advanced scientific concepts specially in the area of condensed matter physics, optics, electromagnetism and electronics, and quantum mechanics which host advanced phenomena and applications. \cite{asadchy2020tutorial, caloz2018nonreciprocity}

In order to achieve non-reciprocity, the most important task in electromagnetics and optical systems is to break time-reversal symmetry. The most conventional method to create non-reciprocity is by applying external magnetic fields to ferromagnetic compounds called ferrites, such as Yttrium Iron Garnet (YIG) and materials composed of iron oxides and other elements (Al,Co,Mn,Ni)\cite{gurevich1996magnetization,lax1962microwave, caloz2018nonreciprocity}. However, due to technological advancement and enormous possibilities of this field, nonreciprocity has been investigated by various researchers over the last decade where major focus was invested to achieve nonreciprocity without the presence of static magnet \cite{ULMKodera,Swadesh, Kodera1, caloz2018nonreciprocity, meta}.

One of the key performance indicator is polarization rotation and this can be defined as the rotation of the orientation of the plane of polarization about the microwave/optical axis of linearly polarized wave as it moves through certain materials. Material properties and the bias condition plays a pivotal role in achieving specific types of rotation.   

\begin{figure}[htbp]
    \centering
    \includegraphics[height=.21\textheight]{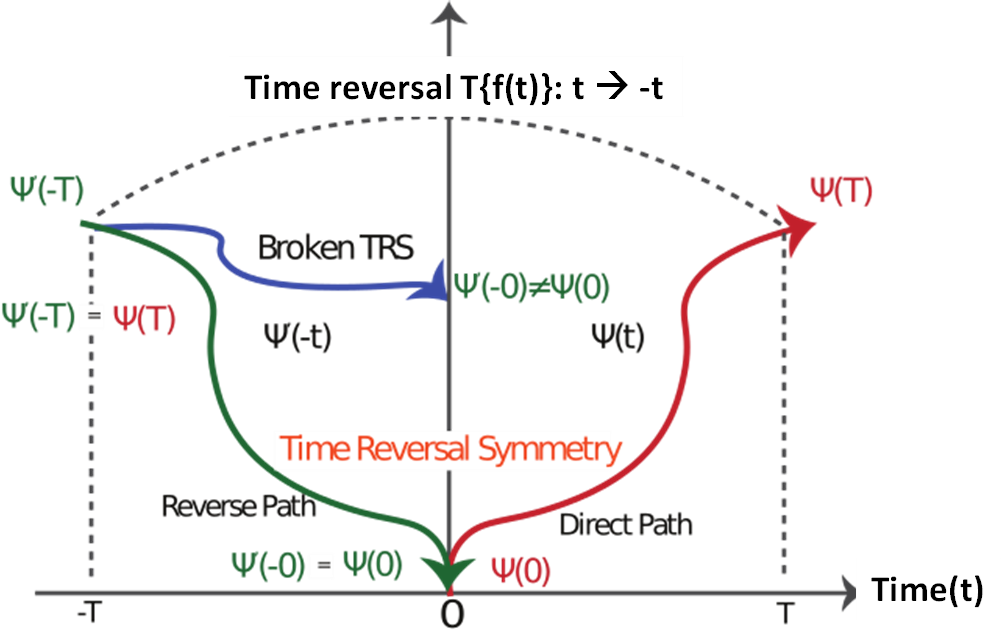}
    %\captionsetup{justification=justified,width=0.4\textwidth}
    \caption[Time-reversal symmetry (TRS) (red and green curves) and
broken time-reversal symmetry (red and blue curves)...]
    {Time-reversal symmetry (TRS) (red and green curves) and
broken time-reversal symmetry (red and blue curves). Reproduced with permission\cite{caloz2018nonreciprocity} Copyright 2018, American Physical Society}
    \label{fig:TRS_def}
  \end{figure}

The concept of electromagnetic reciprocity is closely related to that of the time-reversal symmetry of Maxwell’s equations. Mathematically, from Fig. \ref{fig:TRS_def}, time reversal symmetry can be defined as
\vspace{-0.5cm}
\begin{equation}
\mathit{T}(t)=t^{'}=-t \rightleftharpoons  \mathit{T}:t\rightarrow t^{'}=-t,
\label{eq:TRS-def1}
\end{equation}
where in Eq. \ref{eq:TRS-def1}, time reversal is defined by the operator T. A process like Fig. \ref{fig:TRS_def} can be represented as

\begin{equation}
\mathit{T}({\psi(t)})=\psi^{'}(t^{'})=\psi^{'}(-t).
\label{eq:TRS-def2}
\end{equation}

Based on the above Fig. \ref{fig:TRS_def}, Eq. \ref{eq:TRS-def1} and Eq. \ref{eq:TRS-def2}, if the system remains the same or changes under the reversal of time,
a system can be defined as time reversal symmetric or asymmetric from 
\begin{equation}
T\begin{Bmatrix}
\psi (t)
\end{Bmatrix}=\psi ^{'}(-t)\begin{Bmatrix}
=\\ 
\neq
\end{Bmatrix}\psi ^{}(t).
\label{eq:TRS-def3}
\end{equation}

The fundamental concept of time reversal symmetry and asymmetry provides the required foundation for reciprocity and nonreciprocity\cite{caloz2018nonreciprocity, asadchy2020tutorial}. Besides, the basic laws of physics are classically invariant under time-reversal symmetry and can be intuitively visualized by the Fig. \ref{fig:TRS_def}. 

\subsection{Time Reversal Symmetry Breaking}
In presence of external or internal bias (in the non-linear case), time-reversal symmetry can be broken.
The physical quantities involved in the laws of physics denoted as \emph f(t) may be either time-reversal symmetric or time-reversal antisymmetric and can be represented as
\begin{equation}
T\left \{ f(t) \right \}=f^{'}(t^{'})=f^{'}(-t^{'})=\pm f(-t),
\label{eq:TRS-def4}
\end{equation}
where “$+$” corresponds to time-reversal symmetry, or even time-reversal parity, and “$-$” corresponds to time-reversal antisymmetry, or odd time-reversal parity \cite{caloz2018nonreciprocity}. Equation \ref{eq:TRS-def4} depicts that all physical quantities are either even or odd under time reversal \cite{caloz2018nonreciprocity,asadchy2020tutorial}.

An electromagnetic wave consists of a coupled oscillating electric field and magnetic field which are always perpendicular to each other and by convention, the "polarization" of electromagnetic waves refers to the direction of the electric field. In linear polarization, the fields oscillate in a single direction. In circular or elliptical polarization, the fields rotate at a constant rate in a plane as the wave travels. This wave can be decomposed into two orthogonal polarization states (right handed and left handed). 
%propagation in anisotropic media that are circularly birefringent. 
A linear-polarized wave can be decomposed into two circularly-polarized waves, one right-hand circularly polarized (RHCP), and another left-hand circularly-polarized (LHCP). When a circular polarized wave propagates in a birefringent media, RHCP and LHCP waves propagate with different speeds. After propagating, when viewed as a linearly-polarized wave (recombining the RHCP and LHCP components), this difference causes the linear-polarized wave to have it's polarization plane rotate as they reflect from the anisotropic media which is known as
%-an effect known as natural optical or 
optical or Kerr rotation. The type of the media those experience these specific phenomenon are circularly birefringent and 
%exhibit these properties. 
examples are sugar solutions, proteins,  nucleic acids, amino acids, lipids, DNA, vitamins, hormones, and natural substances \cite{orfanidis2016electromagnetic}. On the other hand, linearly birefringent materials can also be used to change one polarization into another, such as changing linear into circular. The polarization of the field keeps changing as it propagates and relative phase between x-y plane that is introduced by this propagation is called retardence. The relative phase in any reciprocal system can be written as \cite{chen1983theory,orfanidis2016electromagnetic,balanis2012advanced}
Polarization rotation varies on various medium and is directly related to material properties, system condition. For example polarization rotation in chiral, wire grid polarizer, ferrite, plasma, birefringent, and gyrotropic medium are different. If a linearly polarized wave travels forward through a nonreciprocal medium by a distance $l$, gets reflected, and travels
back to the starting point, the polarization rotation angle depend on the medium properties, distance and wavelength which can be represented as \cite{orfanidis2016electromagnetic, Swadesh}.

\begin{equation}
\phi = (n_1-n_2)\frac{2\pi l}{\lambda},
\end{equation}
where, $l$ is the medium length, $\lambda$ is the wavelength , $n_1$ and $n_2$ are refractive indices defined as $n_1$ = $\sqrt\frac{\varepsilon_1}{\varepsilon_0}$ and $n_2$ = $\sqrt\frac{\varepsilon_2}{\varepsilon_0}$.
%Therefore, the nature of the polarization of the field keeps changing as it propagates. The 

Electromagnetic chirality is closely related to the mirror asymmetry, polarization rotation, and magneto-electric coupling \cite{caloz2019electromagnetic}. The rotation through chiral media can be represented as
\begin{equation}
\phi = \frac{1}{2}(\emph k_\circlearrowright- \emph k_\circlearrowleft)l,
\label{eq:chiral_3}
\end{equation}
where $k_\circlearrowright$ = $n_\circlearrowright \emph k_0$ = $\omega(\sqrt{\mu \varepsilon} + \chi)$ and $k_\circlearrowleft$ = $n_\circlearrowleft \emph k_0$ = $\omega(\sqrt{\mu \varepsilon} - \chi)$. Here, n defines refractive index of the medium and $K_0$ represents the free space wave number.

\begin{figure}[htbp]
    \centering
    \includegraphics[width=0.8\columnwidth]{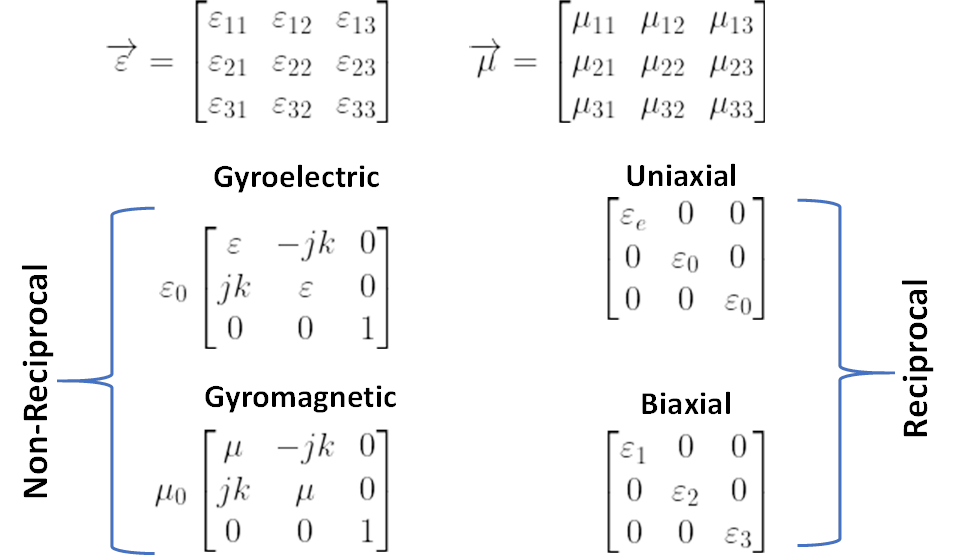}
    %\captionsetup{justification=justified,width=0.4\textwidth}
    %\caption[Biased ferrite material in normal incidence...]
    \caption{Matrix representation of various media and classification of Non-reciprocal and Reciprocal medium}
    \label{fig:Equations}
  \end{figure}

Ferromagnetic materials exhibit non-reciprocal gyrotropic response when they are biased with a static magnetic field. Besides, we have discussed in an earlier section that the permeability tensor of the ferrite has off-diagonal components, therefore, gyrotropic response of the magnetized ferrite is evident. This can be easily understood by considering the constitutive relation $\mathbf{B}=\overrightarrow{\mathbf{\mu }}\cdot \mathbf{H}$, where the x component of magnetic flux density has contribution both from x and y component of the magnetic field intensity. In addition, from the reciprocity theorem \cite{1450781}, the reciprocal birefringent media has to satisfy the following conditions
\begin{equation}
\overrightarrow{\mathbf{\varepsilon }}=\overrightarrow{\mathbf{\varepsilon }}^{T},
\label{eq:Reciprocity_cond_1}
\end{equation}
\begin{equation}
\overrightarrow{\mathbf{\mu }}=\overrightarrow{\mathbf{\mu }}^{T},
\label{eq:Reciprocity_cond_2}
\end{equation}
\begin{equation}
\overrightarrow{\mathbf{\zeta }}=-\overrightarrow{\mathbf{\xi }}^{T},
\label{eq:Reciprocity_cond_3}
\end{equation}
where $(.)^T$ is the Hermitian operator, $\overrightarrow{\mathbf{\zeta }}$ and $\overrightarrow{\mathbf{\xi }}$ are electric-magnetic and magneto-electric coupling tensors, respectively. Figure \ref{fig:Equations} shows the fundamental mathematical representation of various medium and classify them as reciprocal and nonreciprocal. Therefore, it can be clearly seen that, although, wave propagating through various media such as chiral, wire-grid, uniaxial or biaxial experience polarization rotation, those media do not exhibit non-reciprocal gyrotropic properties, hence, can not be used in non-reciprocal devices. 
\vspace{0.1cm}
In summary, the key things that necessitate non-reciprocal gyrotropic response for devices are intrinsic properties of material and external bias to break time reversal symmetry. Next section will be focused on developing analytical model for the measurement of polarization rotation (Faraday and Kerr) in a combination of isotropic and anisotropic media.
Non-reciprocal gyrotropy is a response of certain materials (e.g. Ferrites, Plasma) such that the medium rotates the polarization plane of an electromagnetic wave in presence of magnetic bias by a different amount other than the negative of the rotation angle when the medium is excited from the receiving port with same transmitted field pattern (time reversed case), as can be seen in the rotation angle in this media is called the Faraday rotation ($\theta_F$). 

Figure \ref{fig:Equations} represents the matrix form of various reciprocal and nonreciprocal cases. Nonreciprocity requires asymmetric electric permittivity and magnetic permeability. The properties can be represented as Gyroelectric (${\varepsilon} \neq \mathbf{{\varepsilon}^{T}}$)  and Gyromagnetic (${\mu} \neq \mathbf{{\mu}^{T}}$), respectively where in Fig.\ref{fig:Equations}, k in the off-diagonal terms is the Gyrotropic parameters. Nonreciprocal Gyrotropic and Gyromagnetic materials exhibit off diagonal tensor elements, whereas, uniaxial or biaxial material don't have any off diagonal elements, hence reciprocal. In addition of the off diagonal tensor components, nonreciprocal material exhibits cross polarized phase and magnitude difference \cite{Swadesh, meta}. However, there are specific situations such as anisotropic media with optical axes not aligned with the coordinate systems also have off-diagonal elements yet are reciprocal or where the systems with symmetric tensors ${\varepsilon} = \mathbf{{\varepsilon}^{T}}$ or ${\mu} = \mathbf{{\mu}^{T}}$ with nonzero off-diagonal elements exhibit polarization rotation (optical activity), however, in a  reciprocal way \cite{krasnok}.

\section{Mathematical Modelling of Faraday and Kerr Rotation For Biased Ferrite Material}

\vspace{0.1cm}
  \begin{figure}[htbp]
    \centering
    \includegraphics[width=1\columnwidth]{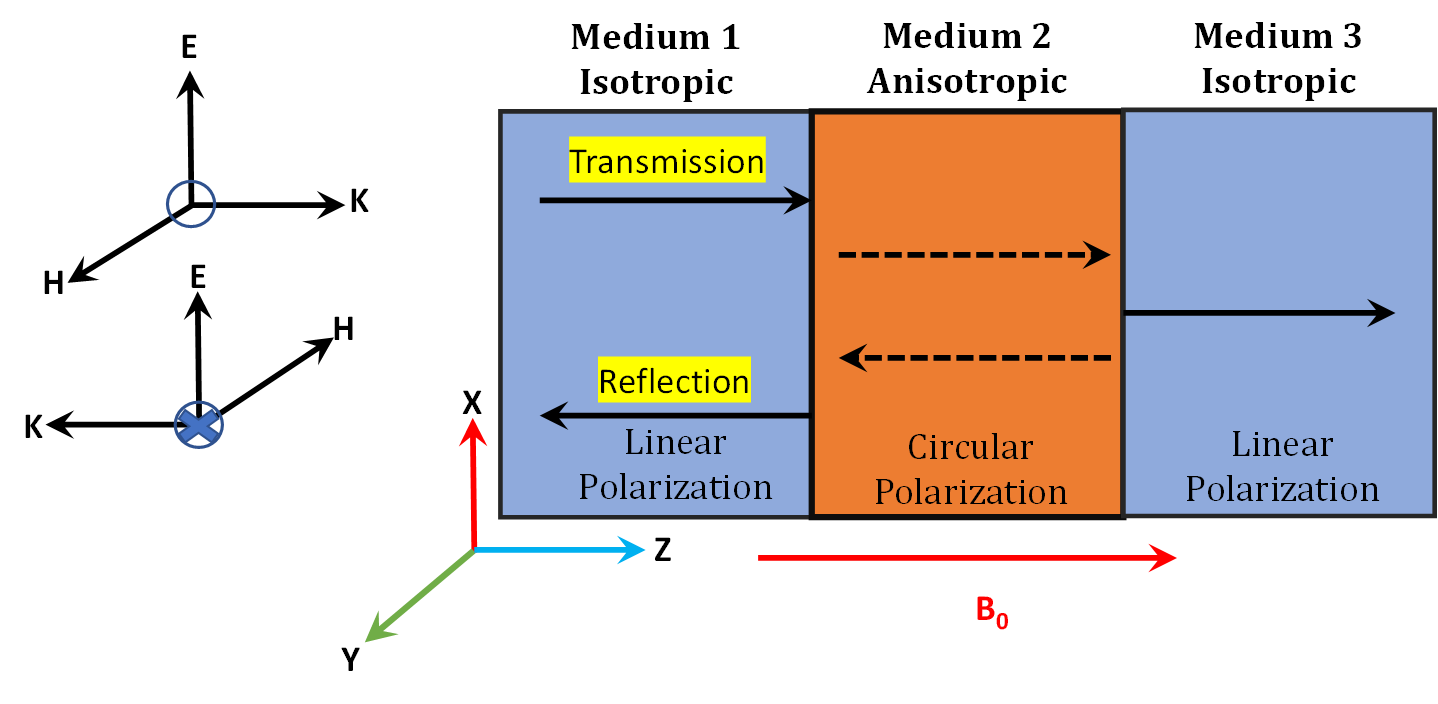}
    %\captionsetup{justification=justified,width=0.4\textwidth}
    \caption[Biased ferrite material in normal incidence...]
    {Biased anisotropic media in normal incidence }
    \label{fig:appendix_fig}
  \end{figure}

\begin{figure}[htbp]
    \centering
    \includegraphics[width=1\columnwidth]{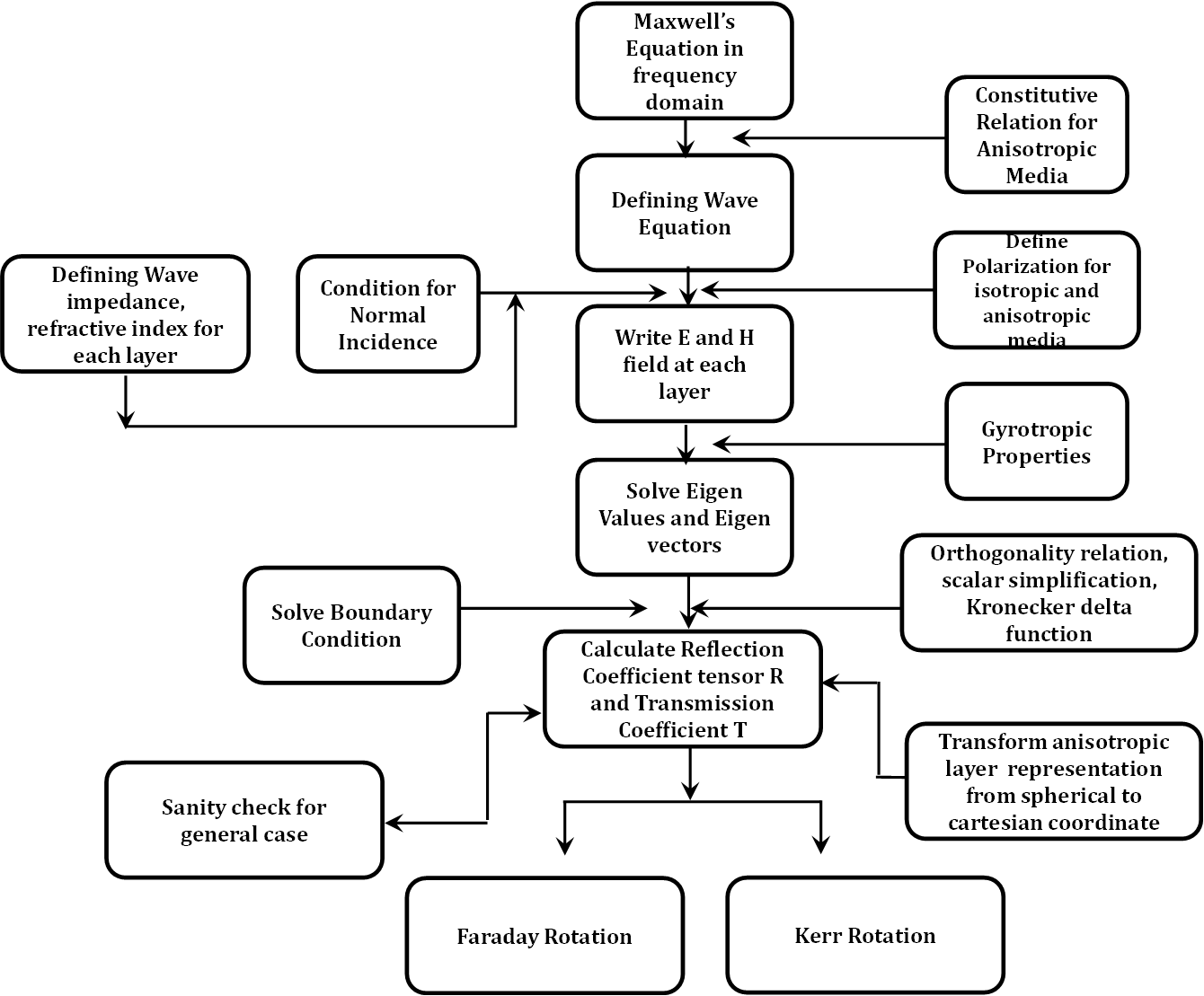}
    %\captionsetup{justification=justified,width=0.4\textwidth}
    \caption[Biased ferrite material in normal incidence...]
    {Process flow of the analytical modeling }
    \label{fig:flowchart}
  \end{figure}

In this section, we will develop a robust mathematical modelling to represent polarization rotation for a combination of isotropic and anisotropic media shown in Fig. \ref{fig:appendix_fig}. Wave propagation direction and axis information are
shown in the figure. Figure \ref{fig:flowchart} shows the step-by-step procedure of our mathematical modeling approaches. 

\vspace{0.1cm}

\subsection{Electric Field}
Maxwell's equations in frequency domain in the \emph p-th medium (p = 1,2,3) can be written as%

\begin{equation}
i\mathbf{k\times E}_{p}=i\omega \mu _{0}\mathbf{\vec{\mu}}^{(p)}\cdot \mathbf{%
H}_{p},
\label{Eq:eq3}
\end{equation}
\begin{equation}
i\mathbf{k\times H}_{p}=-i\omega \varepsilon _{0}\mathbf{\vec{\varepsilon}}%
^{(p)}\cdot \mathbf{E}_{p}.
\label{Eq:eq4}
\end{equation}
where $\mathbf{\vec{\mu}}^{(p)}$ and $\mathbf{\vec{\varepsilon}}^{(p)}$ are
tensors, $\mathbf{E_p}$ and $\mathbf{H_p}$ are the corresponding electric and magnetic field in the pth medium and $\mathbf{k}$ is the wave vector.

We introduce \textbf{$\vec{K}$ } as the tensor that satisfies the below relation%
\begin{equation}
\mathbf{k}\times \mathbf{E}_{p}=\mathbf{\vec{K}\cdot E}_{p},
\label{Eq:eq9}
\end{equation}%
\begin{equation}
\mathbf{k}\times \mathbf{H}_{p}=\mathbf{\vec{K}\cdot H}_{p}.
\label{Eq:eq10}
\end{equation}

where matrix form of tensor \textbf{$\vec{K}$} for pth medium can be represented as

\begin{equation}
\mathbf{\vec{K}}_{p}=\left[ 
\begin{array}{ccc}
0 & -k_{z}^{(p)} & k_{y}^{(p)} \\ 
k_{z}^{(p)} & 0 & -k_{x}^{(p)} \\ 
-k_{y}^{(p)} & k_{x}^{(p)} & 0%
\end{array}%
\right].
\label{Eq:eq11}
\end{equation}

%From above Eq. \ref{Eq:eq1} - \ref{Eq:eq11}, this can be written as%
Therefore, wave equation can be represented in a combined form as
\begin{equation}
\left\{ \frac{\omega ^{2}}{c^{2}}(\mathbf{\vec{K}}_{p}^{-1}\cdot \mathbf{%
\vec{\mu}}^{(p)}\cdot \mathbf{\vec{K}}_{p}^{-1}\cdot \mathbf{\vec{\varepsilon%
}}^{(p)})+\mathbf{I}\right\} \cdot \mathbf{E}_{p}=0.
\label{Eq:eq13}
\end{equation}

For normal incidence, $k_{x}^{(p)}=k_{y}^{(p)}=0,$ therefore, the Eq. \ref{Eq:eq13} can be written in eigen value form as follows

\begin{equation}
\left[ 
\begin{array}{cc}
M_{x,x}^{(p)} & M_{x,y}^{(p)} \\ 
M_{y,x}^{(p)} & M_{y,y}^{(p)}%
\end{array}%
\right] \cdot \left[ 
\begin{array}{c}
E_{p,x} \\ 
E_{p,y}%
\end{array}%
\right] =-\left\{ n^{(p)}\right\} ^{-2}\left[ 
\begin{array}{c}
E_{p,x} \\ 
E_{p,y}%
\end{array}%
\right],
\label{Eq:eq18}
\end{equation}

where refractive index, $n^{(p)}$ is defined as%
\begin{equation}
n^{(p)}\equiv \frac{ck_{z}^{(p)}}{\omega },
\label{Eq:eq15}
\end{equation}%
and,%

\begin{align}
%\begin{equation}
M_{x,x}^{(p)}=\mu _{y,x}^{p}\varepsilon _{y,x}^{p}-\varepsilon _{x,x}^{p}\mu
_{y,y}^{p},
%\end{equation}%
%\begin{equation}
M_{x,y}^{(p)}=\mu _{y,x}^{p}\varepsilon _{y,y}^{p}-\mu _{y,y}^{p}\varepsilon
_{x,y}^{p},
%\end{equation}%
%\begin{equation}
\\
M_{y,x}^{(p)}=\mu _{y,x}^{p}\varepsilon _{y,y}^{p}-\mu _{y,y}^{p}\varepsilon
_{x,y}^{p},
%\end{equation}%
%\begin{equation}
M_{y,y}^{(p)}=\mu _{x,y}^{p}\varepsilon _{x,y}^{p}-\mu _{x,x}^{p}\varepsilon
_{y,y}^{p}.
%\end{equation}
\end{align}
%Therefore, Eq. \ref{Eq:eq16} can be written as%
The following requirement is required to be satisfied%
\begin{equation}
Det\left( \left\{ n^{(p)}\right\} ^{-2}\left[ 
\begin{array}{cc}
M_{x,x}^{(p)} & M_{x,y}^{(p)} \\ 
M_{y,x}^{(p)} & M_{y,y}^{(p)}%
\end{array}%
\right] +\left[ 
\begin{array}{cc}
1 & 0 \\ 
0 & 1%
\end{array}%
\right] \right) = 0,
\label{Eq:eq19}
\end{equation}
yields the following two eigen-values%
%\begin{align}
%\begin{equation} \label{Eq:eq20}
\begin{multline*}
%\usepackage{breqn}
%\begin{split}
%\begin{dmath}
%\begin{multiline}
\left\{ n_{j}^{(p)}\right\} ^{2} =
\zeta _{j}^{(p)} = \\
-\frac{1}{2}\left( M_{x,x}^{(p)}+M_{y,y}^{(p)}
+s_{j}\sqrt{%
(M_{x,x}^{(p)}-M_{y,y}^{(p)})^{2}+4M_{x,y}^{(p)}M_{y,x}^{(p)}}\right)
%\end{multiline}
%\end{dmath}
%\end{split}
%\end{equation}
%\end{align}
\end{multline*}

and the following eigen-vectors%
\begin{equation}
\mathbf{V}_{p,j}=\frac{1}{\sqrt{1+\left\vert \frac{\zeta
_{j}^{(p)}-M_{x,x}^{(p)}}{M_{x,y}^{(p)}}\right\vert ^{2}}}\left[ 
\begin{array}{c}
1 \\ 
\frac{\zeta _{j}^{(p)}-M_{x,x}^{(p)}}{M_{x,y}^{(p)}}%
\end{array}%
\right] ;\text{ }j=1,2,
\label{Eq:eq21}
\end{equation}
with the coefficient $s_{j}$ being given as%
\begin{equation*}
s_{1}=1,s_{2}=-1
\end{equation*}

We, hereafter, assume medium 1 and 3 are isotropic (gyrotropic with no off-diagonal element) i.e.
\begin{equation}
\varepsilon _{x,x}^{(p)}=\varepsilon _{y,y}^{(p)}\text{ , }\varepsilon
_{x,y}^{(p)}=-\varepsilon _{y,x}^{(p)}=0\text{ ; }p=1,3
\end{equation}%
\begin{equation}
\mu _{x,x}^{(p)}=\mu _{y,y}^{(p)}\text{ , }\mu _{x,y}^{(p)}=-\mu
_{y,x}^{(p)}=0\text{ ; }p=1,3,
\label{Eq:eq22}
\end{equation}
and we also assume that medium 2 is gyrotropic%
\begin{equation}
\varepsilon _{x,x}^{(2)}=\varepsilon _{y,y}^{(2)}\text{, }\varepsilon
_{x,y}^{(2)}=-\varepsilon _{y,x}^{(2)}=-i\varepsilon _{g}\neq 0,
\label{Eq:eq23}
\end{equation}%
\begin{equation}
\mu _{x,x}^{(2)}=\mu _{y,y}^{(2)}\text{, }\mu _{x,y}^{(2)}=-\mu
_{y,x}^{(2)}=-i\mu _{g}\neq 0.
\label{Eq:eq24}
\end{equation}

Therefore, 
\begin{equation}
M_{x,x}^{(p)}=M_{y,y}^{(p)}\text{, }M_{x,y}^{(p)}=-M_{y,x}^{(p)},
\label{Eq:eq25}
\end{equation}
and so the eigen values reduce to%
\begin{equation}
\left\{ n_{j}^{(p)}\right\} ^{2}=\zeta
_{j}^{(p)}=-(M_{x,x}^{(p)}+is_{j}M_{x,y}^{(p)}).
\label{Eq:eq26}
\end{equation}

The refractive index of each medium is given as%
\begin{equation}
n^{(1)}=\sqrt{\mu _{x,x}^{(1)}\varepsilon _{x,x}^{(1)}},
\label{Eq:eq27}
\end{equation}%
\begin{eqnarray}
%\begin{split}
n_{j}^{(2)} &=&\sqrt{-(M_{x,x}^{(2)}+is_{j}M_{x,y}^{(2)})} \\
&=&\sqrt{\left( \varepsilon _{x,x}^{(p)}+is_{j}\varepsilon
_{x,y}^{(p)}\right) \left( \mu _{x,x}^{(p)}+is_{j}\mu _{x,y}^{(p)}\right) },
%\end{split}
\end{eqnarray}%
\begin{equation}
n^{(3)}=\sqrt{\mu _{x,x}^{(3)}\varepsilon _{x,x}^{(3)}},
\label{Eq:eq28}
\end{equation}

where, n defines the refractive index, superscript and subscript describe the medium of wave propagation and LHCP/RHCP in anistropic media respectively. Furthermore, the eigen-vector corresponding to gyrotropic medium take the following simplified form%
\begin{equation}
\mathbf{V}_{p,j}=\frac{\mathbf{\hat{e}}_{x}+is_{j}\mathbf{\hat{e}}_{y}}{%
\sqrt{2}}=\frac{1}{\sqrt{2}}\left[ 
\begin{array}{c}
1 \\ 
is_{j}%
\end{array},%
\right] \text{, }j=\circlearrowright ,\circlearrowleft
\label{Eq:eq29}
\end{equation}

The eigen-vectors for gyrotropic medium correspond to left handed
circularly polarized (LHCP) wave and right handed circularly polarized
(RHCP) wave. Here, we adopt, $j=$ $\circlearrowright ,\circlearrowleft $ labels
instead of $j=1,2.$

Since the eigen-vectors given by Eq. \ref{Eq:eq29} from a complete set in
two dimensional space, the electromagnetic field in all three media can be
expressed as a linear combination of these eigen-vectors as they represents the amplitudes of the propagating waves inside the medium. In appendix A, E field is shown at various medium where medium 1 and medium 3 are isotropic medium and medium 2 is gyrotropic. Therefore during propagation, EM wave will exhibit phase difference, different refractive indices and different impedances for RHCP and LHCP waves.

\subsection{Magnetic Field}

By following the similar approach described in the previous section, wave equation can be represented as

\begin{equation}
\begin{split}
-i\omega \mu _{0}\mathbf{\vec{\mu}}^{(p)}\cdot \sum_{j=\circlearrowright
,\circlearrowleft }H_{p,j}^{\pm }\mathbf{\hat{E}}_{j}\exp (\mp \frac{%
iz\omega }{c}n_{j}^{(p)})\\
=\sum_{j=\circlearrowright ,\circlearrowleft }(\mp 
\frac{i\omega }{c})E_{p,j}^{\pm }n_{j}^{(p)}[\mathbf{\hat{e}}_{z}\mathbf{%
\times \hat{E}}_{j}]\exp (\mp \frac{iz\omega }{c}n_{j}^{(p)}),
\label{Eq:eq36}
\end{split}
\end{equation}
where we introduced $\mathbf{\hat{E}}_{\circlearrowright }=\frac{\mathbf{\hat{e}}_{x}-i\mathbf{%
\hat{e}}_{y}}{\sqrt{2}}$, $\mathbf{\hat{E}}_{\circlearrowleft }=\frac{\mathbf{\hat{e}}_{x}+i\mathbf{%
\hat{e}}_{y}}{\sqrt{2}}$ and $\mathbf{\hat{e}}_{z}\times \mathbf{\hat{E}}_{j}=-is_{j}\mathbf{\hat{E}}_{j}$ as identities.%

Here, $s_{\circlearrowright }=1$ and $s_{\circlearrowleft }=-1$

and Eq. \ref{Eq:eq36} can be reduced to 
\begin{equation}
\mathbf{\vec{\mu}}^{(p)}\cdot \mathbf{\hat{E}}_{j} = (\mu _{x,x}^{(p)}+is_{j}\mu _{x,y}^{(p)})\mathbf{\hat{E}}_{j}
\end{equation}
Similarly magnetic field in pth medium can be represented in a combined form%
\begin{equation}
H_{p,j}^{\pm }=\mp is_{j}\frac{E_{p,j}^{\pm }}{Z_{p,j}},
\end{equation}
where $Z_{p,j}^{(p)}$ is the impedance of the p-th medium experienced by the
j-th plane wave $(j=\circlearrowright ,\circlearrowleft)$ 
\begin{equation}
Z_{j}^{(p)}=Z_{0}\sqrt{\frac{\mu _{x,x}^{(p)}+is_{j}\mu _{x,y}^{(p)}}{%
\varepsilon _{x,x}^{(p)}+is_{j}\varepsilon _{x,y}^{(p)}}}.
\label{Eq:eq40}
\end{equation}
with, $Z_{0}=\sqrt{\frac{\mu _{0}}{\varepsilon _{0}}}=377 \Omega ,$ impedance of the free space.

\subsection{Reflection and Transmission Coefficient}

In previous section we have described electric and magnetic field and based on the mathematical modelling developed for E and H field reflection and transmission coefficient will be described here. Medium 1 (Isotropic) to medium 2 (Gyrotropic). 

Incident and reflected wave can be represented as for the medium 1 %
\begin{equation}
\mathbf{E}_{p}^{\pm}=(E_{p,\circlearrowright }^{\pm}\mathbf{\hat{E}}%
_{\circlearrowright }+E_{p,\circlearrowleft }^{\pm}\mathbf{\hat{E}}%
_{\circlearrowleft })\exp (\mp\frac{i\omega zn^{(p)}}{c}).
\label{Eq:eq41}
\end{equation}
where p = 1-3 defines the layer number.

where superscript "$+$" or "$-$" defines incident or reflected wave.

By following the continuity equation and boundary condition at various medium interface, the magnetic field corresponding to the incident, reflected, and transmitted fields are given as%
\begin{equation}
H_{p}^{\pm }=\sum_{j=\circlearrowright ,\circlearrowleft }H_{p,j}^{\pm }%
\mathbf{\hat{E}}_{j}\exp (\mp \frac{iz\omega }{c}n_{j}^{(2)}),
\end{equation}

where $Z_{r}^{(p)}$ denotes the relative impedance of the \emph p-th isotropic
medium (p = 1,3) and is defined as follows%
\begin{equation}
Z_{r}^{(p)}=Z_{0}\sqrt{\frac{\mu _{x,x}^{(p)}}{\varepsilon _{x,x}^{(p)}}}
\end{equation}

Similarly, magnetic field can be represented  for every medium as

\begin{equation}
\mathbf{H}_{p}^{\pm}={\mp}is_{\circlearrowright }\frac{E_{p,\circlearrowright }^{\pm}%
}{Z_{r}^{(p)}}\mathbf{\hat{E}}_{\circlearrowright }\exp (\mp\frac{iz\omega }{c}%
n_{\circlearrowright }^{(p)}),
\end{equation}%
where p = 1-3 defines the layer number.

Medium 3 is assumed as no reflection for simplicity. Therefore, there will be no reflected wave for that medium.

Due to the orthogonality of the left-hand and right hand eigen vector, i.e., 
$\mathbf{\hat{E}}_{n}\cdot \mathbf{\hat{E}}_{m}^{\ast }=\delta _{n,m}$ for $%
n,m=\circlearrowright ,\circlearrowleft ,$ the preceding vector equations
can be expressed in terms of the scalar equations and by following some algebraic calculation leads us to the following expression where E field can be represented in terms of medium 1 and medium 3 at z = d interface as below.
%\begin{multline*}
\begin{equation}
%\begin{split}
{E_{3,j}^{+}=\frac{4E_{1,j}^{\pm}}{L + M}}
%\end{split}
\end{equation}
%\end{multline*}

where
\begin{align*}
    L = (1+\frac{Z_{j}^{(2)}}{Z_{r}^{(3)}}\pm \frac{Z_{r}^{(1)}%
}{Z_{r}^{(3)}}\pm \frac{Z_{r}^{(1)}}{Z_{j}^{(2)}})\exp (\frac{i\omega
dn_{j}^{(2)}}{c}) \\
M = (1-\frac{Z_{j}^{(2)}}{Z_{r}^{(3)}}\pm \frac{Z_{r}^{(1)}}{%
Z_{r}^{3}}\mp \frac{Z_{r}^{(1)}}{Z_{j}^{(2)}})\exp (-\frac{i\omega
dn_{j}^{(2)}}{c})
\end{align*}

\bigskip

Comparing the preceding equation with the definition of transmission tensor yields,%
\begin{equation}
t_{n,m}^{c}=\frac{4\delta _{n,m}}{U+V}
\end{equation}

where
\begin{equation}
U=(1+\frac{Z_{j}^{(2)}}{Z_{r}^{(3)}}+ \frac{Z_{r}^{(1)}
}{Z_{r}^{(3)}}+ \frac{Z_{r}^{(1)}}{Z_{j}^{(2)}})\exp (\frac{i\omega
dn_{j}^{(2)}}{c})
\end{equation}
\begin{equation}
V = (1-\frac{Z_{j}^{(2)}}{Z_{r}^{(3)}}+ \frac{Z_{r}^{(1)}}{%
Z_{r}^{3}}- \frac{Z_{r}^{(1)}}{Z_{j}^{(2)}})\exp (-\frac{i\omega
dn_{j}^{(2)}}{c})
\end{equation}

with $\delta _{n,m\text{ }}$and $t_{n,m}^{c}$ respectively denoting the
Kronecker delta function and components of the transmission tensor in
LHCP-RHCP basis set, i.e.%

\begin{equation}
\mathbf{\vec{t}=}\sum_{n=\circlearrowright ,\circlearrowleft
}\sum_{m=\circlearrowright ,\circlearrowleft }t_{n,m}^{c}\mathbf{\hat{E}}_{n}%
\mathbf{\hat{E}}_{m}^{\ast }\\
=\left[ 
\begin{array}{cc}
t_{\circlearrowright ,\circlearrowright }^{c} & t_{\circlearrowright
,\circlearrowleft }^{c} \\ 
t_{\circlearrowleft ,\circlearrowright }^{c} & t_{\circlearrowleft
,\circlearrowleft }^{c}%
\end{array}%
\right].
\end{equation}

Therefore, transmission tensor is represented as%
\begin{equation}
\mathbf{\vec{t}=}\sum_{m=\circlearrowright ,\circlearrowleft }\frac{4}{U+V}\mathbf{\hat{E}}_{m}%
\mathbf{\hat{E}}_{m}^{\ast}
\end{equation}
The components of transmission tensor can be represented in cartesian coordinates as%
\begin{equation}
\begin{split}
\mathbf{\vec{t}=}\sum_{n=x,y}\sum_{m=x,y}t_{n,n}\mathbf{\hat{e}}_{n}\mathbf{%
\hat{e}}_{m}
\equiv \left[ 
\begin{array}{cc}
t_{x,x} & t_{x,y} \\ 
t_{y,x} & t_{y,y}%
\end{array}%
\right].
\end{split}
\end{equation}
In order to obtain the components of the transmission tensor in Cartesian
coordinates, $\mathbf{\hat{E}}_{\circlearrowleft }$ and $\mathbf{\hat{E}}%
_{\circlearrowright }$ should be substituted with their Cartesian representation%
\begin{equation}
\mathbf{\vec{t}}=t_{\circlearrowright ,\circlearrowright }^{c}(\frac{\mathbf{%
\hat{e}}_{x}-i\mathbf{\hat{e}}_{y}}{\sqrt{2}})(\frac{\mathbf{\hat{e}}_{x}+i%
\mathbf{\hat{e}}_{y}}{\sqrt{2}})+t_{\circlearrowleft ,\circlearrowleft }^{c}(%
\frac{\mathbf{\hat{e}}_{x}+i\mathbf{\hat{e}}_{y}}{\sqrt{2}})(\frac{\mathbf{%
\hat{e}}_{x}-i\mathbf{\hat{e}}_{y}}{\sqrt{2}}).
\end{equation}

Therefore, the components of the transmission tensor in Cartesian coordinates are given as%
\begin{equation}
t_{x,x}=t_{y,y}=\sum_{m=\circlearrowright ,\circlearrowleft }\frac{2}{U+V}
\label{eq:txx}
\end{equation}

\begin{equation}
t_{y,x}=-t_{x,y}=i\sum_{m=\circlearrowright ,\circlearrowleft }\frac{2s_{m}}{U+V}
\label{eq:txy}
\end{equation}

\bigskip
Similarly, reflection tensor can be obtained as with $r_{n,m}^{c}$ denoting the components of the reflection tensor in
LHCP-RHCP basis set as%

\begin{equation}
\mathbf{\vec{r}}=\sum_{n=\circlearrowright ,\circlearrowleft
}\sum_{m=\circlearrowright ,\circlearrowleft }r_{n,m}^{c}\mathbf{\hat{E}}_{n}%
\mathbf{\hat{E}}_{m}^{\ast }\equiv \left[
\begin{array}{cc}
r_{\circlearrowright ,\circlearrowright }^{c} & r_{\circlearrowright
,\circlearrowleft }^{c} \\ 
r_{\circlearrowleft ,\circlearrowright }^{c} & r_{\circlearrowleft
,\circlearrowleft }^{c}%
\end{array}%
\right],
\end{equation}

The components of reflection tensor can be represented in Cartesian
coordinates as
\begin{equation}
\begin{split}
\vec{r}=\sum_{n=x,y}\sum_{m=x,y}r_{n,m}\mathbf{\hat{e}}_{n}\mathbf{\hat{e}}%
_{m}
\equiv \left[ 
\begin{array}{cc}
r_{x,x} & r_{x,y} \\ 
r_{y,x} & r_{y,y}%
\end{array}%
\right].
\end{split}
\end{equation}

The transmission and reflection tensor can be expressed in terms of $\theta _{m}$ which is the polarization angle of the linearly polarized plane wave at $z=0$
\begin{equation}
\theta _{m}=\frac{\omega dn_{m}^{(2)}}{c}=2\pi \frac{n_{m}^{(2)}d}{\lambda
_{0}}
\end{equation}

After some calculation steps and algebraic steps, Transmission tensor components for a thin slab, i.e., $\theta
_{\circlearrowright \circlearrowleft }$%
\begin{equation}
t_{x,x}=t_{y,y}=2\sum_{m=\circlearrowright
,\circlearrowleft}(1+\frac{%
Z_{r}^{(1)}}{Z_{r}^{(3)}})-i(\frac{Z_{m}^{(2)}}{Z_{r}^{(3)}}+\frac{%
Z_{r}^{(1)}}{Z_{m}^{(2)}})\theta _{m},
\end{equation}
\begin{equation}
t_{y,x}=-t_{x,y}=2\sum_{m=\circlearrowright ,\circlearrowleft }(\frac{%
Z_{m}^{(2)}}{Z_{r}^{(3)}}+\frac{Z_{r}^{(1)}}{Z_{m}^{(2)}})s_{m}\theta _{m}.
\end{equation}
Writing impedance of the media in terms of permitivity and permeability
\begin{eqnarray}
\frac{Z_{m}^{(2)}}{Z_{r}^{(3)}} \simeq &\sqrt{\frac{\varepsilon _{x,x}^{(3)}}{\mu _{x,x}^{(3)}}}\sqrt{\frac{%
\mu _{x,x}^{(2)}}{\varepsilon _{x,x}^{(2)}}}\left[ 1+\frac{is_{m}}{2}\left( 
\frac{\mu _{x,y}^{(2)}}{\mu _{x,x}^{(2)}}-\frac{\varepsilon _{x,y}^{(2)}}{%
\varepsilon _{x,x}^{(2)}}\right) \right],
\end{eqnarray}%
\begin{eqnarray}
\frac{Z_{r}^{(1)}}{Z_{m}^{(2)}} \simeq &\sqrt{\frac{%
\varepsilon _{x,x}^{(2)}}{\mu _{x,x}^{(2)}}}\sqrt{\frac{\mu _{x,x}^{(1)}}{\varepsilon _{x,x}^{(1)}}}\left[ 1-\frac{is_{m}}{2}\left( 
\frac{\mu _{x,y}^{(2)}}{\mu _{x,x}^{(2)}}-\frac{\varepsilon _{x,y}^{(2)}}{%
\varepsilon _{x,x}^{(2)}}\right) \right],
\end{eqnarray}%
\begin{equation}
\begin{split}
\theta _{m}=2\pi \frac{d}{\lambda _{0}%
}\sqrt{\varepsilon _{x,x}^{(2)}\mu _{x,x}^{(2)}}\left[ 1+\frac{is_{m}}{2}%
\left( \frac{\mu _{x,y}^{(2)}}{\mu _{x,x}^{(2)}}+\frac{\varepsilon
_{x,y}^{(2)}}{\varepsilon _{x,x}^{(2)}}\right) \right].
\end{split}
\end{equation}

\subsection{Faraday and Kerr Rotation}
In previous section we have described the mathematical representation of fields in different layers, medium classification in terms of impedance, permeability, permittivity, and transmission and reflection tensor. In the following section we will develop the analytical model for the polarization rotation in an anisotropic media, known as Faraday and Kerr rotation. Following some trigonometric mathematical derivation, Faraday rotation can be written in below form
\begin{eqnarray}
\theta _{F}
&\simeq &2\pi \frac{n_{r}^{(2)}d}{\lambda _{0}}\frac{\frac{\mu _{g}}{\mu
_{xx}^{(2)}}Z_{r}^{(2)}Z_{r}^{(3)}+\frac{\varepsilon _{g}}{\varepsilon
_{xx}^{(2)}}\frac{Z_{r}^{(1)}}{Z_{r}^{(2)}}}{1+Z_{r}^{(1)}Z_{r}^{(3)}}.
\end{eqnarray}

As a validation of our mathematical modelling this can be easily shown from above derivation that for layer thickness $d=0$ and for $Z_{r}^{(1)}=Z_{r}^{(3)}=Z_{\circlearrowright
}^{(2)}=Z_{\circlearrowleft }^{(2)}$ and $n_{\circlearrowleft
}^{(2)}=n_{\circlearrowright }^{(2)}=n_{r}^{(2)},$ the expressions for reflection and transmission follows the conventional reflection and transmission forms.%

From above equations Faraday rotation can be expressed into more robust form considering the angle of the incident wave.
\begin{equation}
\begin{split}
\theta _{F}
=\tan ^{-1}\left( \frac{t_{y,x}[\mathbf{E}_{1}^{+}(z=0)\cdot \mathbf{\hat{e%
}}_{x}]+t_{y,y}[\mathbf{E}_{1}^{+}(z=0)\cdot \mathbf{\hat{e}}_{y}]}{t_{x,x}[%
\mathbf{E}_{1}^{+}(z=0)\cdot \mathbf{\hat{e}}_{x}]+t_{x,y}[\mathbf{E}%
_{1}^{+}(z=0)\cdot \mathbf{\hat{e}}_{y}]}\right)\\
-\theta _{pol},
\end{split}
\end{equation}
By following some trigonometric calculation 

\begin{equation}
\theta _{F}=\tan ^{-1}\left( \frac{t_{y,x}\cos \theta _{pol}+t_{y,y}\sin
\theta _{pol}}{t_{x,x}\cos \theta _{pol}+t_{x,y}\sin \theta _{pol}}\right)
-\theta _{pol},
\label{eq:Faraday}
\end{equation}
By replacing transmission and reflection tensor Faraday rotation can be represented in a very comprehensive form.
\begin{equation}
\theta _{F}=\tan ^{-1}\left( \frac{i\cos \theta
_{pol}\sum_{m_{\circlearrowright ,\circlearrowleft}}A+\sin \theta
_{pol}\sum_{m_{\circlearrowright ,\circlearrowleft} }B}{\cos \theta
_{pol}\sum_{m_{\circlearrowright ,\circlearrowleft} }B-i\sin \theta
_{pol}\sum_{m_{\circlearrowright ,\circlearrowleft} }A}\right) -\theta _{pol},
\label{eq:76}
\end{equation}
where
\begin{equation}
A=\frac{s_{m}}{(1+\frac{Z_{r}^{(1)}}{Z_{r}^{(3)}})\cos \theta _{m}+i(\frac{%
Z_{m}^{(2)}}{Z_{r}^{(3)}}+\frac{Z_{r}^{(1)}}{Z_{m}^{(2)}})\sin \theta _{m}},
\end{equation}%
\begin{equation}
B=\frac{A}{s_m},
\end{equation}%

Therefore this mathematical formulation is applicable for any arbitrary polarization angle.
\bigskip

When the bias is perpendicular to the slab, Faraday rotation is independent of on the polarization of the incident field. i.e.,%
\begin{equation*}
\frac{\partial \theta _{F}}{\partial \theta _{pol}}=0
\end{equation*}
Therefore, the expression Eq. \ref{eq:76} for Faraday rotation can be simplified for the condition ($\theta _{pol} = 0$)%

\begin{equation}
\theta _{F}=\tan ^{-1}\left( \frac{i\sum_{m=
\circlearrowright ,\circlearrowleft }A}{\sum_{m=\circlearrowright
,\circlearrowleft }B}\right).
\end{equation}

Similarly, Kerr Rotation can be simplified to

\begin{equation}
\theta _{K}=\tan ^{-1}\left( \frac{i\cos \theta
_{pol}\sum_{m=\circlearrowright ,\circlearrowleft }E+\sin \theta
_{pol}\sum_{m=\circlearrowright ,\circlearrowleft }F}{\cos \theta
_{pol}\sum_{m=\circlearrowright ,\circlearrowleft }F-i\sin \theta
_{pol}\sum_{m=\circlearrowright ,\circlearrowleft }E}\right),
\end{equation}
where 
\begin{equation}
E=s_{m}\frac{(1-\frac{Z_{r}^{(1)}}{Z_{r}^{(3)}})\cos \theta _{m}+i(\frac{%
Z_{m}^{(2)}}{Z_{r}^{(3)}}-\frac{Z_{r}^{(1)}}{Z_{m}^{(2)}})\sin \theta _{m}}{%
(1+\frac{Z_{r}^{(1)}}{Z_{r}^{(3)}})\cos \theta _{m}+i(\frac{Z_{m}^{(2)}}{%
Z_{r}^{(3)}}+\frac{Z_{r}^{(1)}}{Z_{m}^{(2)}})\sin \theta _{m}},
\end{equation}%
\begin{equation}
F=\frac{E}{s_m},
\end{equation}%

When the bias is perpendicular to the slab, Kerr rotation is independent of on the polarization of the incident field. i.e.,%
\begin{equation}
\frac{\partial \theta _{K}}{\partial \theta _{pol}}=0.
\end{equation}
Therefore, the expression for Kerr rotation can be simplified for the condition $\theta _{pol}$ = 0 as follow %

\begin{equation}
\theta _{K}=\tan ^{-1}\left( \frac{i\sum_{m_
{\circlearrowright ,\circlearrowleft} }E}{\sum_{m_{\circlearrowright ,\circlearrowleft} }F
}\right).
\end{equation}

The above described mathematical modelling can be used for numerical computation of both Faraday and Kerr rotation and can be extended in any number of gyrotropic, non-gyrotropic layers. Based on the medium impedance and thickness of the slab, both Faraday and Kerr rotation can be calculated.

\section{Analytical Model Validation}
In this section we will crosscheck/validate our analytical model and generalize for known conditions.
For the condition of a very thin slab when d = 0, the co-polarization transmission component at equation \ref{eq:txx} reduces to

\begin{equation}
\begin{split}
t_{x,x}=t_{y,y}=\frac{Z^{(3)}_{r}}{Z^{(3)}_{r}+ Z^{(1)}_{r}}\sum_{m = \circlearrowleft, \circlearrowright}1 =  \frac{2Z^{(3)}_{r}}{Z^{(3)}_{r}+ Z^{(1)}_{r}}
\end{split}
\label{eq:new_txx}
\end{equation}

Similarly, cross polarization component at \ref{eq:txy} reduces to

\begin{equation}
\begin{split}
t_{x,y}=-t_{y,x}=\frac{iZ^{(3)}_{r}}{Z^{(3)}_{r}+ Z^{(1)}_{r}}\sum_{m =\circlearrowleft, \circlearrowright}s_m = 0
\end{split}
\label{eq:new_txy}
\end{equation}
where $s_m$ = $\pm1$ for LHCP and RHCP wave. This also satisfies the general condition that there will be no cross polarization compoenent for conventional cases.

Similarly co polarized reflection components can be shown as 
\begin{equation}
\begin{split}
r_{x,x}=r_{y,y}=
\frac{Z^{(3)}_{r}-Z^{(1)}_{r}}{Z^{(3)}_{r}+ Z^{(1)}_{r}},
\end{split}
\label{eq:new_rxx}
\end{equation}
and cross polarized reflection component can be shown as
\begin{equation}
\begin{split}
r_{x,y}=-r_{y,x}=\frac{i}{2}
\frac{Z^{(3)}_{r}-Z^{(1)}_{r}}{Z^{(3)}_{r}+ Z^{(1)}_{r}}\sum_{m = \circlearrowleft, \circlearrowright}s_m = 0,
\end{split}
\label{eq:new_rxy}
\end{equation}

In another condition when the normalized impedance  at every layer are follows the condition $Z^{1}_{r}$ = $Z^{3}_{r}$ = $Z^{2}_{\circlearrowleft}$ = $Z^{2}_{\circlearrowright}$ and the refractive indices are $n^{2}_{\circlearrowleft}$ = $n^{2}_{\circlearrowright}$ = $n^{2}_{r}$, the expression for reflection and transmission reduces to 
\begin{equation}
\begin{split}
r_{x,y}=-r_{y,x}= 0,
\end{split}
\label{eq:new_rxy2}
\end{equation}
\begin{equation}
\begin{split}
r_{x,x}=r_{y,y}= 0,
\end{split}
\label{eq:new_rxx2}
\end{equation}
\begin{equation}
\begin{split}
t_{x,y}=-t_{y,x}= 0,
\end{split}
\label{eq:new_txy2}
\end{equation}
\begin{equation}
\begin{split}
t_{x,x}=t_{y,y}= exp(-\frac{i\omega dn^{(2)}_{r}}{c}),
\end{split}
\label{eq:new_txx2}
\end{equation}
which is inline with the known matched conditions.

Furthermore, to validate the robustness of the model we consider a simple middle layer, i.e. no gyrotropy behaviour in medium. After some trigonometric and algebraic simplification the co-polarized reflection coefficient can be simplified into below conventional form

\begin{equation}
r_{x,x}= r_{y,y}= \Gamma_{in}=\frac{\Gamma_{12}+\Gamma_{23}exp(-\frac{2i\omega dn^{(2)}}{c})}{1 + \Gamma_{12}\Gamma_{23}exp(-\frac{2i\omega dn^{(2)}}{c})}
\label{eq: simplifiedform},
\end{equation}
where $\Gamma$ is the reflection coefficient defines the ratio between reflected and incident wave.

In the next step we will try to reproduce an already published artificial metasurface developed by Kodera et al. \cite{Kodera5} and use our proposed mathematical model to evaluate the Faraday and Kerr rotation. In order to design and reproduce the nonreciprocal metasurface, following the procedure developed by our previous work \cite{meta}, here we computed the floquet mode analysis and based on Jones calculus which has been discussed in detail in another article currently under review, we mapped our proposed model to floquet mode/S parameter in simulation environment and calculate Faraday and Kerr rotation. The overall performance matches closely with the results reported in the reference article. Therefore, we can claim that this mathematical model is can be helpful to extract Faraday or Kerr rotation from nonreciprocal environment using the transmission or reflection matrices.

\vspace{0.1cm}
  \begin{figure}[htbp]
    \centering
    \includegraphics[width=1\columnwidth]{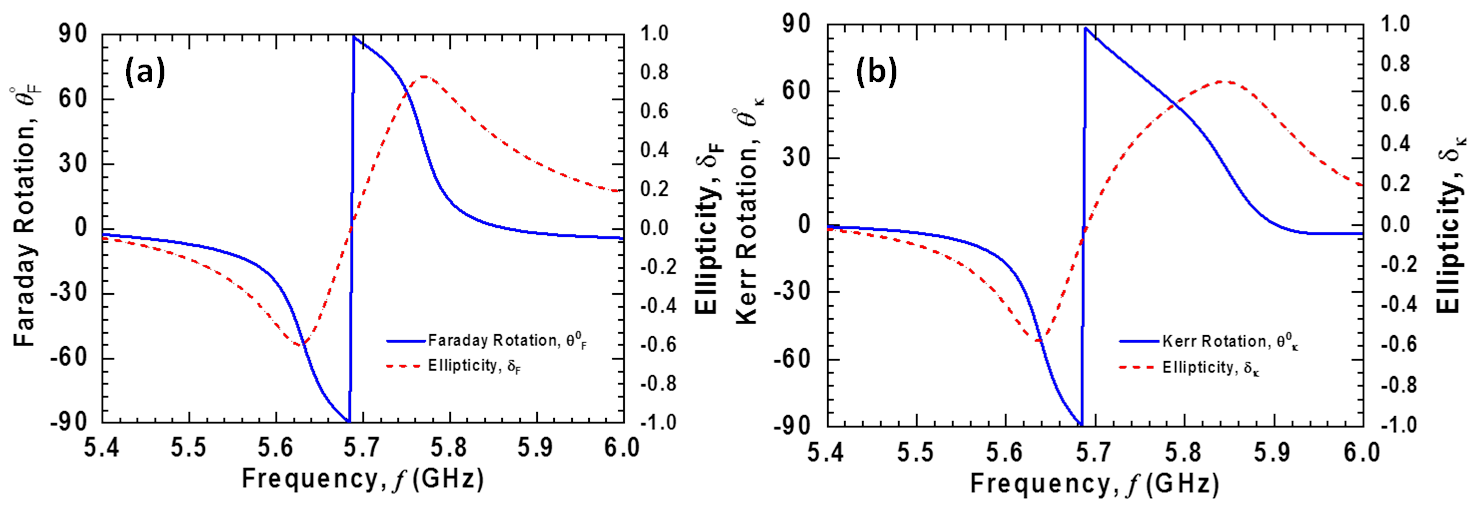}
    %\captionsetup{justification=justified,width=0.4\textwidth}
    \caption[Biased ferrite material in normal incidence...]
    {(a) Faraday and (b) Kerr rotation of metasurface structure using our proposed analytical model }
    \label{fig:FR_KR}
  \end{figure}

\section{Conclusion}
We have presented an analytical approach of modeling polarization rotation in an anisotropic media. Starting from Maxwell's equation, we have developed transmission, reflection tensor leading to polarization rotation leading to analytical model validation for known conditions and compare/contrast with previously reported results. The mathematical model has been explained in terms of three layer media, however, this approaches/outcome should be valid for arbitrary number of layers. We envision that this work will be helpful in the nonreciprocal research and add a building block.

\section{Appendix}

\bigskip
E field and H field in various medium: 
In medium 1 (isotropic), i.e., 
$-\infty <z<0,$ the forward and backward propagating waves are described by%
\begin{equation}
\mathbf{E}_{1}^{\pm }=(E_{1,\circlearrowright }^{\pm }\mathbf{\hat{E}}%
_{\circlearrowright }+E_{1,\circlearrowleft }^{\pm }\mathbf{\hat{E}}%
_{\circlearrowleft })\exp \left( \mp \frac{i\omega zn^{(1)}}{c}\right),
\label{Eq:eq30}
\end{equation}
and in medium 2 (anisotropic), i.e., $0<z<d,$ the forward and backward propagating waves are
described by%
\begin{equation}
\mathbf{E}_{2}^{\pm }=E_{2,\circlearrowright }^{\pm }\mathbf{\hat{E}}%
_{\circlearrowright }\exp \left( \mp \frac{i\omega zn_{\circlearrowright }}{c%
}\right) +E_{2,\circlearrowleft }^{\pm }\mathbf{\hat{E}}_{\circlearrowleft
}\exp \left( \mp \frac{i\omega zn_{\circlearrowleft }}{c}\right),
\label{Eq:eq31}
\end{equation}
and in medium 3 (isotropic), i.e., $d<z<\infty $ the forward propagating waves are described by%
\begin{equation}
\mathbf{E}_{3}^{+}=(E_{3,\circlearrowright }^{+}\mathbf{\hat{E}}%
_{\circlearrowright }+E_{3,\circlearrowleft }^{+}\mathbf{\hat{E}}%
_{\circlearrowleft })\exp \left( -\frac{i\omega (z-d)n^{(3)}}{c}\right).
\label{Eq:eq32}
\end{equation}

\maketitle{}
\bibliographystyle{IEEEtran}
\bibliography{bibliography.bib}
\end{document}